\documentclass[conference]{IEEEtran}
\usepackage{graphicx,booktabs,amsmath,cite}
\usepackage{xcolor}
\usepackage{xurl}
\usepackage[most]{tcolorbox}
\newtcolorbox{rqbox}{colback=orange!18,colframe=black!55,boxrule=0.5pt,arc=2pt,left=4pt,right=4pt,top=3pt,bottom=3pt}
\newtcolorbox{observationbox}{colback=orange!6,colframe=orange!65!black,boxrule=0.5pt,arc=2pt,left=4pt,right=4pt,top=3pt,bottom=3pt}

\usepackage[hidelinks]{hyperref}

\begin{document}

\title{5G CATS: 5G-Based Cryptographic Authorization Against Trafficking of
Semiconductor Devices}
\author{
  \IEEEauthorblockN{Sreevikram Rajaraman, Derek Ladd, Nicholas Heinrich-Barna, Srinivas Shakkottai, and Jeyavijayan Rajendran}
  \IEEEauthorblockA{Department of Electrical and Computer Engineering\\
    Texas A\&M University, College Station, TX, USA\\
    }
}
\maketitle

\begin{abstract} 
Enforcing export controls on AI accelerators requires verifying where a chip operates after deployment. Round-trip time measurements can bound a chip's distance from a trusted landmark, but any jitter in honest replies forces the verifier to loosen its timeout, giving an attacker slack to relay traffic over fiber from a restricted region. Prior approaches either cross the untrusted host stack or rely on noisy power-line channels that inflate this slack.

We present 5G CATS, a location certification system that pairs a co-packaged 5G radio, cryptographic engine, and hardware interlock on the GPU package with 5G Ultra-Reliable Low-Latency Communication (URLLC) base stations. Terminating the radio link inside the package removes the host software, PCIe fabric, and baseboard management controller from the verification path. We map an authenticated challenge-response protocol onto 5G's slot structure, replace permanent key erasure with short-lived execution leases, and specify a cross-check of the chip's reported serving cell against the carrier's own attachment records. In ns-3 and 5G-LENA simulations of an unloaded cell, our URLLC-based radio profile narrows the 99.9th percentile round-trip spread to $2.60$ ms, more than ten times narrower than default 5G. A tracer rejects every round of a fiber relay from 300 km, against 2,000 km for default 5G. Against a 3,000 km attacker, a tracer stays safe at up to 1,410 km from the chip, against at most 619 km for PowerCATS's published model, while default 5G is never safe.
\end{abstract}
\section{Introduction}
\subsection{Motivation}
Frontier large language models increasingly exhibit capabilities with national security implications~\cite{mythos_cve}. Training and serving them requires large clusters of advanced AI accelerators and GPUs, which is a market led by a small number of U.S. chip designers. Since October 2022, the U.S. has imposed successive export controls to keep these chips out of restricted destinations. However, these controls only govern sales rather than verifying where chips end up~\cite{csis_export_controls}.

The gap is being exploited and the clearest case starts in California. In August 2025, federal prosecutors in Los Angeles charged two Chinese nationals from Pasadena and El Monte with shipping export-controlled Nvidia GPUs to China through freight forwarders in Singapore and Malaysia. The alleged exports ran from October 2022 to July 2025~\cite{doj2025_arrests}. On 1 October 2026, the owner of Earthmade Computer, a California server business, was arrested on an indictment alleging that more than \$300 million of servers with A100, H100, RTX 4090 and RTX 5090 GPUs were routed through Malaysia and Singapore to China using false end-user paperwork~\cite{register2026_california_arrests}~\cite{doj2026_300m}. Nvidia GPUs worth an estimated \$1B reportedly reached restricted destinations in the three months to July 2025 alone~\cite{ft2025smuggling}. In each case investigators reconstructed the diversion from shipping and payment records, years or months later. The chips themselves reported nothing.

\subsection{Limitations of Existing Approaches}
The simplest option is \textit{\textbf{software telemetry}}, such as NVIDIA's opt-in fleet-management service~\cite{nvidia_ngc_opt_in}, in which an agent installed by the operator reports the chip's status. Since the agent runs under the operator's control, the operator is also able to disable or remove it. \textit{\textbf{Delay-based schemes}} avoid this dependence by having the chip sign replies to landmark servers and using the round-trip time to bound its distance from each landmark~\cite{cset_tracking_chips}. However, each reply passes through the host OS, drivers, NIC, and Internet routing, all of which add jitter, and the verifier's timeout must be set loosely enough to accommodate it. \textit{\textbf{Identity schemes}}, including supply-chain tracking, physically unclonable functions, and hardware fingerprinting, can distinguish individual chips but do not provide information about their location~\cite{blockchain_provenance,Kong2014PUFatt}.

\textit{\textbf{Hardware-enforced mechanisms}} place an interlock on the chip itself~\cite{petrie2025technicaloptionsflexiblehardwareenabled}. PowerCATS~\cite{powercats} combines such an interlock with timing measurements over the power line, and in principle can both locate a chip and disable it. However, power lines are poorly suited to precise timing. Signals transmitted over them are subject to impulsive noise, attenuation, and impedance changes as loads switch on and off~\cite{plc_limitations}. Its delay model is closed-form and assumes zero queuing with a static bit error rate, and its own limitations section names both assumptions as open.

\subsection{Challenges and Contributions}
The cellular network is a natural alternative to the power line since it already gives the chip a carrier-operated, timed link to a known location. However, a simple port of an existing hardware-enforced scheme to default 5G fails. By default 5G we mean the enhanced mobile broadband (eMBB) service that most public cells run. Timing-based location is only as good as its worst honest round trip: the verifier sets its threshold above the slowest honest reply, and every millisecond of slack gives an attacker about 100 km of fiber relay to hide behind. We simulated this naive approach and identified three challenges, which lead to our research question:

\begin{rqbox}
\textbf{Research Question.} Can a chip's location be certified over 5G tightly enough to bound a fiber relay, without disabling honest chips and without trusting the chip's own report of where it is?
\end{rqbox}

We introduce 5G CATS, a set of mechanisms that make a cellular round trip tight, attested and safe to act on. Each mechanism answers one challenge. 

\noindent \textbf{Challenge 1: Scheduled delay.} A cellular round trip is set by the MAC scheduler, not by the physical propagation, so a closed-form delay bound is wrong in shape, not just in constants. On default 5G, honest round trips span about 27 to 57\,ms, and a 750\,km fiber relay adds less than 1\,ms to the median.

\noindent \textbf{Solution 1: Tail-aware timing on URLLC.} We add a fifth requirement, a bounded honest tail, and replace the zero-queue closed form with a percentile threshold and a hideable-relay estimate. 5G CATS runs the location check on 5G's Ultra-Reliable Low-Latency Communication (URLLC) service, which in simulation cuts the variation to 2.60\,ms, and lets the tracer reject every round of a 300\,km relay. Control message traces attribute 2.27\,ms of a 3.65\,ms median to the uplink scheduling-request and grant handshake. Configured-grant uplink and mini-slots, which the simulator lacks, would remove that handshake; we have not measured their effect. 

\noindent \textbf{Challenge 2: False Rejections.} Any percentile threshold rejects a small share of honest rounds. An interlock that erases its key after a single failed round therefore disables honest chips, even at the 99.9th percentile.

\noindent \textbf{Solution 2: Execution leases.} 5G CATS replaces key erasure with short-lived execution leases, so a missed round suspends computation until the next passed round without destroying key material. A sequential test over the chip's misses then separates honest chips from relayed ones.

\noindent \textbf{Challenge 3: An imitable landmark.} The location claim rests on a carrier-operated cell, which an attacker can imitate with a radio of their own.

\noindent \textbf{Solution 3: Carrier-attested landmark binding.} The tracee includes the identity of its serving cell in each reply, and the tracer verifies it against the carrier's own attachment records and the base station's timing-advance measurements, reducing a rogue base station to a relay attack. We specify this binding but have not yet simulated it.

\begin{figure}[t]
\includegraphics[width=\columnwidth]{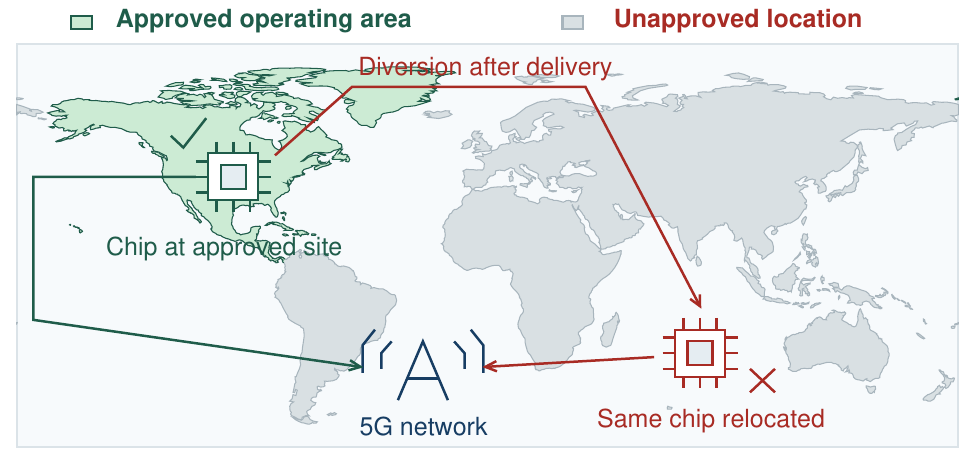}
\caption{Chip diversion and proposed 5G CATS location verification using 5G. Regions represent an example deployment policy.}
\label{fig:overview}
\end{figure}

\section{Background}
\label{sec:background}
\subsection{Threat Model and Requirements}
We consider an adversarial operator who owns the chip and wants to run it outside an approved region. The adversary controls the host software, network, power, and physical environment, including the 5G backhaul and the radio medium between the chip and the 5G base station (gNB). They may eavesdrop on, delay, drop, replay, or relay any transmission, and may operate a base station of their own. We assume that the GPU package is tamper-resistant and that the private 5G cell at each approved site is operated by a licensed carrier or neutral host rather than by the chip's owner.

Following PowerCATS~\cite{powercats}, a traceability scheme must provide (R1) location certification, so that a verifier (the tracer) can determine whether the chip (the tracee) is in an approved region, (R2) a cryptographic interlock that acts on a failed certification, (R3) periodic recertification, so that the tracee cannot be moved after deployment, and (R4) resistance to replay and relay. We add (R5) a bounded honest tail: the gap between the tracer's acceptance threshold and the fastest honest round trip must be small and stable, because it sets how much fiber relay an attacker can hide.

\subsection{Low Latency 5G - URLLC}
5G URLLC is a core 5G service category that targets 1 ms latency at 99.999\% reliability for applications requiring high stability and real-time capabilities like remote surgery, automated controls and smart grids~\cite{urllc_paper}. Compared to prior cellular generations, 5G introduces a flexible air interface and edge-deployable core functions, allowing traffic to be processed at the cell site rather than routed through distant servers. The default 5G service category, enhanced mobile broadband (eMBB), optimizes for bandwidth and throughput, serving applications such as video streaming and large data transfers. URLLC instead prioritizes latency and reliability by changing four mechanisms. It schedules in mini-slots of 2, 4, or 7 symbols rather than full 14-symbol slots, which cuts the wait for a slot boundary. It replaces the scheduling-request handshake with configured grants that pre-allocate uplink resources. Its packets can preempt resources already assigned to bulk traffic. It also uses conservative modulation and uplink repetition instead of relying on Hybrid Automatic Repeat Request (HARQ) retransmissions, each of which adds a variable round trip.

Slot length is $1/2^\mu$\,ms for numerology $\mu$. A time-division duplex (TDD) pattern such as DDDSU at $\mu = 2$ offers an uplink slot every 1.25\,ms against one per 10 ms frame in a typical default 5G pattern. The gNB also advances each device's uplink by its measured round-trip propagation (timing advance), so the network holds a radio-distance measurement for every attached device. Each cell broadcasts a physical cell identity (PCI), one of 1,008 values reused across a network, and a globally unique NR cell global identity.

\section{Approach}
\label{sec:approach}

\subsection{System Overview}
\label{sec:overview}
5G CATS has three parties. The \textit{tracee} is a GPU subject to export control. The \textit{tracer} is a verifier, either a regulator or an entity acting on its behalf, that decides whether a tracee is in an approved region. The third is the operator of the site's private 5G cell, a licensed carrier or neutral host, which provides the serving 5G base station (gNB) and the fiber backhaul to the tracer. The tracer knows the gNB's location and treats it as the landmark where it sits at the operator's edge next to the user plane function (UPF). We assume that the tracee's 5G baseband, RF front end, cryptographic engine and interlock are co-packaged with the GPU. Current packaging supports this: baseband and RF can share a system-in-package~\cite{wells2023aip}, and die-to-die standards connect chiplets within one package~\cite{ucie_spec}. The certification path therefore never crosses the host OS, the PCIe fabric or the baseboard management controller, and the measured round trip ends at the silicon being verified. Any other placement leaves a stretchable link between radio and key that becomes slack in the distance bound.

\noindent \textbf{Deployment.} We target data centers and enterprise sites that host export-controlled accelerators, not consumer GPUs in homes. Each approved site has a private 5G cell run by a licensed carrier or neutral host, never by the chip's owner, and provisioned with the URLLC configuration of Section~\ref{sec:solutions}. One cell may serve hundreds to thousands of chips. The tracer staggers their challenges, and the cell reserves configured-grant uplink resources and downlink capacity for the CATS flow, sized to the number of chips and the round rate, so that recertification rarely needs to preempt other traffic. Our simulation has one device per cell, so this sizing is an assumption that we have not evaluated.

\subsection{Adopted from PowerCATS}
We adopt the PowerCATS elements that do not depend on the medium. Requirements R1-R4 state what a scheme must achieve rather than how, so they carry over largely unchanged. We add R5 because keeping the tail short for any authentication method is critical to effectively recertifying the chip. The four-phase protocol carries over, with its messages sent on one 5G data bearer. Per-round replay protection by a packet ID under AES-GCM~\cite{AESGCM} needs neither a synchronized clock nor on-chip randomness. The on-chip security perimeter is retained so that the chip itself produces each reply and not the host. Time bounded acceptance is retained because it is the only check that detects a relay forwarding valid ciphertext unchanged~\cite{brands1993}.

We modify two elements, because a cellular link or our evaluation showed the original to be insufficient. The acceptance threshold becomes the measured 99.9th percentile of the honest round trip, because scheduled delay has a tail that the closed-form bound omits. The key erasure is replaced by execution leases, because erasure locks honest chips on false rejections.

\subsection{Location Certification Protocol}
\label{sec:certification-protocol}
\noindent \textbf{Provisioning.} Each tracee is a 5G subscriber, and the tracer binds the chip's device identity to its subscription. The CATS session runs end-to-end above 5G's own authentication, whose keys the carrier holds~\cite{ts33501}. The chip and tracer run a key exchange using ML-KEM~\cite{nist_mlkem} and each party signs its share with its identity key, and the tracer applies the recertification time bound to set up the exchange. We specify this exchange but simulate X25519 without signatures (Section~\ref{sec:eval}).

\noindent \textbf{Recertification.} Each round is one URLLC exchange. At a randomized time, the tracer sends a nonce over the chip's data session on a delay-critical QoS flow (5QI 82), through the UPF to the gNB which transmits it in the next downlink mini-slot. The chip returns, under AES-GCM, the PID, its identity and the NCGI of its serving cell on its next configured-grant opportunity without a scheduling-request handshake. The tracer accepts the round if the seal, the PID, the cell binding and the round-trip time all verify. Our simulations implement neither the mini-slots nor the configured grant, so our measurements still include the handshake. 

\noindent \textbf{Carrier-attested cell binding.} The chip's claim about its cell is not trusted on its own. The tracer checks the sealed NCGI against the carrier's record of the chip's serving cell, held by the access and mobility management function (AMF) and exposed to authorized parties through the network exposure function (NEF) as described in Section~\ref{sec:solutions}, and checks the gNB's timing-advance measurement collected by the location management function against the cell's radius. The PCI is not used because any radio can broadcast it. We specify this check but do not simulate it.

\subsection{Latency Budget}
\label{sec:latency-budget}
Let $T$ be the acceptance threshold the tracer uses in Section~\ref{sec:certification-protocol}, and $T_\text{min}$ the smallest round-trip time observed from a legitimate chip from the gNB. A legitimate round trip varies from round to round because of scheduling, channel conditions, and queueing. The tracer sets $T$ at the 99.9th percentile of this distribution to admit a $10^{-3}$ false-reject rate. An attacker running a fiber relay from the chip to a legitimate 5G cell in an approved region extends the signal path by $d$ kilometers each way, adding $2d/c_f$ to the measured round trip, where $c_f \approx \tfrac{2}{3}c$ is the speed of light in optical fiber (equivalently, about 10~\textmu s of round-trip delay per kilometer of relay). The attack passes verification if $T_\text{min} + 2d/c_f \leq T$, so the hideable relay distance is
\begin{equation}
d_\text{hide} = \frac{c_f\,(T - T_\text{min})}{2}.
\label{eq:dhide}
\end{equation}

A tight bound comes from pulling $T$ close to $T_\text{min}$. Equation~\ref{eq:dhide} is an empirical estimate and not a security guarantee: $T$ and $T_\text{min}$ come from a finite sample, and it assumes that every relayed round is as fast as the fastest honest one and that relay delay adds directly to the round trip.

\noindent \textbf{Round-trip decomposition.} PowerCATS bounds the round trip as $T_\text{bound} = T_\text{prop} + T_\text{proc} + T_\text{rout} + \epsilon$, with zero queueing~\cite{powercats}. On a 5G cell, the honest round trip is
\begin{equation}
T_\text{RT} = T_\text{CN} + T_\text{RAN} + T_\text{proc},
\label{eq:trt}
\end{equation}
\begin{equation}
T_\text{CN} = T_\text{fiber} + T_\text{hop} + T_\text{UPF}, \quad
T_\text{RAN} = T_\text{pipe} + T_\text{wait} + T_\text{SR} + T_\text{air}.
\label{eq:tcn-tran}
\end{equation}
$T_\text{fiber}$ is propagation between the tracer and the core, $T_\text{hop}$ router forwarding, and $T_\text{UPF}$ the user plane function (77 and 517\,\textmu s at p50 and p99 in software~\cite{mishra_upf}; 8.2 and 16.2\,\textmu s with DPDK~\cite{dpdk_upf}). $T_\text{pipe}$ is the gNB's two-slot scheduling pipeline (about 500\,\textmu s at $\mu = 2$~\cite{5glena,oai}), $T_\text{wait}$ the wait for the next downlink or uplink opportunity (up to 1.25\,ms for DDDSU at $\mu = 2$, 10\,ms for default 5G), $T_\text{SR}$ the scheduling-request and grant handshake, and $T_\text{air}$ transmission and decoding, with the chip's turnaround bounded by the N1 and N2 budgets of TS~38.214~\cite{ts38214} (about 0.36\,ms under capability 2 at $\mu = 2$). $T_\text{proc}$, the chip's AES-GCM, takes about 10\,\textmu s, and timing advance removes radio propagation.

PowerCATS's terms map onto $T_\text{fiber}$, $T_\text{proc}$ and $T_\text{hop} + T_\text{UPF}$, and its allowance $\epsilon$ covers the resending of lost packets. $T_\text{pipe}$, $T_\text{wait}$ and $T_\text{SR}$ have no counterpart in its model, and $T_\text{wait}$ and $T_\text{SR}$ vary between rounds and dominate $T - T_\text{min}$, which we measure 39.28\,ms on default 5G (Section~\ref{sec:eval-honest}). 5G CATS shortens $T_\text{wait}$, and configured grant would remove $T_\text{SR}$.

\subsection{Cryptographic Interlock: execution leases}
\label{sec:eval-interlock}
5G CATS keeps PowerCATS's protected memory which releases the tracee's key only after successful certification. PowerCATS overwrites that key when certification fails, so the device cannot decrypt its boot code on any later boot and a tracee whose response misses the time bound stops receiving challenges and fails recertification. Because any percentile threshold rejects some honest rounds, this rule would eventually be triggered even for honest chips. 5G CATS instead issues a signed lease with an expiry for every passed round. The interlock allows compute only while it holds an unexpired lease, halts it when the lease lapses, and resumes it on the next passed round. Long-term keys are never erased. An on-package watchdog lets the lease lapse if challenges stop arriving, so cutting the radio link also halts compute.

Leases alone would let a short relay that fails only some rounds keep computing between failures. The tracer therefore applies a sequential test to the chip's misses over a window of 1,000 rounds. It only flags the chip with 9 or more misses in this rolling period. If misses are independent at the honest false-reject rate of $10^{-3}$, this raises a false alarm with probability about $10^{-6}$ per window. We leave the lease length and the response to a flag as deployment policy.

\subsection{Meeting the Three Challenges}
\label{sec:solutions}
Each challenge from the introduction is met by a specific set of mechanisms, described here.

\noindent \textbf{Challenge 1: Scheduled delay.} The round trip must be short and stable enough that a relay shows up as extra delay. 5G CATS meets this in four steps. First, the CATS flow runs on a URLLC configuration: 0.25\,ms slows, an uplink opportunity every 1.25\,ms, UR processing capability 2, a $10^{-5}$ block-error target and at most one HARQ retransmission. These settings shrink $T_\text{wait}$ and the retransmission delays, the terms that set $T - T_\text{min}$ (Section~\ref{sec:latency-budget}). Second, configured-grant uplink and mini-slots remove $T_\text{SR}$ and most of the wait for a slow boundary; we specify them, but our simulator lacks them. Third the tracer calibrates $T$ for each cell as 99.9th percentile of honest round trips and adds any known backhaul and router delay, which shifts the distribution without widening it. Fourth, the tracer sends each challenge at a randomized time so that rounds sample the whole TDD frame and an attacker cannot predict when the next challenge arrives. With these steps, the honest spread falls to 2.60\,ms in simulation, and every round of a relay from 300\,km exceeds $T$ (Section~\ref{sec:eval}).

\noindent \textbf{Challenge 2: False rejections.} A miss must cost an honest chip little, while repeated misses must still expose a relay. Each round proceeds as follows:
\begin{enumerate}
    \item The tracer checks the seal, the PID, the cell binding and the round-trip time of the reply. 
    \item If all four pass, the tracer returns a lease signed with its identity key, bound to the chip's identity and round's PID, with an expiry. 
    \item If any check fails, no lease is sent. Compute halts when the current lease expires and resumes at the next passed round.
    \item The tracer records every miss and flags the chip when 9 or more of its last 1,000 rounds miss.
\end{enumerate}

An honest miss, about one in thousand rounds, therefore costs at most one lease period, while a relay that misses often accumulates musses until the test flags it. In our replay, no honest chip is halted for more than one round of flagged from 100\,km (Section~\ref{sec:eval-lease}). The lease length and the round rate are deployment parameters; the evaluation uses one lease per passed round.

\noindent \textbf{Challenge 3: An imitable landmark.} The tracer must know that the chip is attached to an approved cell without trusting the chip's own report. The chip seals the NR cell global identity (NCGI) of its serving cell into every reply. The tracer subscribes through the NEF to the AMF's serving-cell change events for the chip's subscription and keeps the latest authenticated record. For each round it compares the sealed NCGI with that record, and when the cell changes, it also checks the gNB's timing-advance measurements against the cell's radius. Because the comparison uses the cached record and runs after the reply has arrived, it lies outside the measured round trip and never delays a lease. A cell change or a mismatch withholds the next lease until the record shows an approved cell. A rogue cell cannot produce a carrier record for an approved cell, so the attacker must forward the traffic through a real approved gNB, which reduces the attack to the relay that Challenge 1 bounds. We specify this binding but do not simulate it.

\section{Attack Vectors}
\label{sec:attacks}
We analyze three attack classes: replay and forgery, rogue base stations, and fiber relays. Of these, fiber relaying poses the central challenge to distance-based certification because it extends the physical path within the slack that honest latency jitter leaves in the tracer's acceptance threshold, while preserving legitimate authentication.

\subsection{Replay and forgery}
An attacker could capture a response and replay it later, or try to forge a new one without the session key. Both fail because every message carries an AEAD tag over a per-round PID: the AEAD tag catches a forged response, and the advancing PID catches a replayed one.

\subsection{Rogue gNB}
An attacker could set up their own 5G cell in the restricted region, broadcasting a spoofed approved-region cell's ID. The chip would faithfully include that cell ID in its AEAD response, but this does not help on its own, because the tracer verifies the cell ID against the network carrier's own records rather than against the chip's claim alone. The carrier knows which gNB each UE is attached to through authenticated N2 signaling, and a rogue radio has no certificate to register with the carrier under a legitimate cell's identity. For the chip to appear registered at the approved cell, the attacker has to forward its traffic through the actual gNB there, which collapses the attack into the fiber relay case below.

\subsection{Fiber relay}
This is the main attack the system targets. An attacker in a restricted region tunnels the chip's traffic over fiber to a legitimate gNB in an approved region. The extra fiber adds propagation delay to the round trip, and if the delay pushes the observed time above the tracer's threshold, the round is rejected.

\section{Evaluation}
\label{sec:eval}
\subsection{Setup}
We evaluate 5G CATS in simulation against two baselines. The first is PowerCATS~\cite{powercats}, for which we use only the numbers its authors publish. The second is default 5G, the naive approach: the same protocol on an enhanced mobile broadband (eMBB) cell, with a permanent key erasure. In the simulation, 5G CATS means the URLLC radio profile with a 99.9th-percentile threshold. We evaluate execution leases and the sequential test by replaying measured rounds in Q5, and we do not simulate the carrier binding. We ask five core questions:

\begin{itemize}
    \item \textbf{Q1:} How much room does the honest round trip leave an attacker to hide a relay?
    \item \textbf{Q2:} Can a tracer detect an attacker who relays a tracee's traffic from a restricted region?
    \item \textbf{Q3:} How far can a tracer sit from a tracee before an attacker can answer just as fast?
    \item \textbf{Q4:} Does the bound hold as the deployment changes?
    \item \textbf{Q5:} Does a rejected round disable an honest tracee?
\end{itemize}

We compare the two radio profiles in Table~\ref{tab:profiles}. The default 5G cell is a bulk-traffic eMBB configuration: 1\,ms slots with an uplink slot pair every 10\,ms, a 10\% block-error target, up to three HARQ retransmissions, and RLC acknowledged mode. The 5G CATS cell uses 0.25\,ms slots with an uplink slot every 1.25\,ms, a $10^{-5}$ block-error target, and at most one HARQ retransmission. Everything else, including cell realizations, is identical. Each cell serves one tracee 300\,m from the gNB with line of sight and no other traffic, through a software UPF and three IP hops, with the tracer beside the 5G core unless stated. The simulator uses ideal beamforming and ideal RRC signaling, and it implements neither configured-grant uplink nor mini-slots, so in both cells the chip's reply waits for a scheduling request and a grant.

We run 20 cell realizations of 1,000 recertifications each. Each challenge leaves at a randomized interval longer than the longest round trip, so that rounds sample the whole TDD frame and do not lock to it. The protocol follows PowerCATS's message flow with an X25519 key exchange without signatures and a 103-byte recertification authenticated by HMAC. Messages have their real sizes and cryptographic execution is a cycle-count model, but tags are stand-in values. A relay is a fixed fiber delay added to each message on the chip's side of the radio link so the cell sees a normal device.

\begin{table}[t]
\centering
\caption{Comparison of Default 5G and 5G CATS Radio Profiles}
\label{tab:profiles}
\begin{tabular*}{\columnwidth}{@{\extracolsep{\fill}}lll@{}}
\toprule
\textbf{Parameter} & \textbf{Default 5G} & \textbf{5G CATS} \\
\midrule
Numerology, slot & $\mu=0$, 1\,ms & $\mu=2$, 0.25\,ms \\
TDD pattern & DDDDDDDSUU & DDDSU \\
Uplink opportunity & Every 10\,ms & Every 1.25\,ms \\
UE processing & Capability 1 & Capability 2 \\
Block-error target & $10^{-1}$ & $10^{-5}$ \\
HARQ retransmissions & At most 3 & At most 1 \\
Scheduler, 5QI & Round-robin, 9 & QoS-aware, 82 \\
RLC mode & Acknowledged & Unacknowledged \\
\bottomrule
\multicolumn{3}{@{}l}{\footnotesize *All other parameters are identical between profiles.}
\end{tabular*}
\end{table}

\begin{figure}[t]
  \includegraphics[width=\columnwidth]{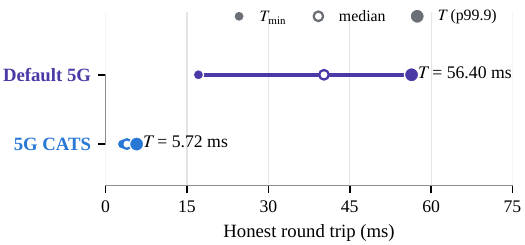}
  \caption{Round trip under default 5G and 5G CATS, tracer at the 5G core; 20 cell realizations $\times$ 1,000 recertifications each. Tail of the distribution, dots mark the threshold $T$ at the 99.9th percentile.}
  \label{fig:embb_vs_urllc}
\end{figure}

\subsection{Q1: PowerCATS vs Default 5G vs 5G CATS}
\label{sec:eval-honest}
A tracer can only reject a tracee whose reply is slower than the slowest honest reply, so any gap between that threshold and the fastest honest reply is room for an attacker to hide a relay. Q1 measures that room for each system.

Fig.~\ref{fig:embb_vs_urllc} and Table~\ref{tab:urllc-embb} compare the three systems. The median round trip is 40.25\,ms under default 5G and 3.92\,ms under 5G CATS. The threshold is 56.40\,ms against 5.72\,ms, and the observed spread $T - T_\text{min}$ is 39.28\,ms against 2.60\,ms. By equation ~\ref{eq:dhide}, these spreads give estimated hidden relays of 3,928\,km and 260\,km. The default 5G value rests on its 2 fastest rounds in 20,000. Measured from the first percentile instead, the spreads are 29.15\,ms and 2.51\,ms, so 5G CATS is more than 10 times narrower on either measure. PowerCATS publishes a closed-form bound with an allowance for lost packets, not a distribution, so it has no comparable spread. Its published curves start from a one-way delay of 5.3 to 7.1\,ms at zero distance, which we double to a round trip of 10.6 to 14.2\,ms, roughly twice the 5G CATS threshold.

Most of the difference between default 5G and 5G CATS is in the uplink. Under default 5G, the uplink takes 35.92\,ms of the 40.25\,ms median. The chip's reply has to wait for an uplink slot, send a scheduling request, receive a grant, and then send its buffer status report and data. With one uplink slot pair per 10\,ms frame, where a challenge lands in the frame decides how long this takes, so honest round trips fall in bands across roughly 27 to 57\,ms. The uplink period of 10\,ms is equivalent to 1,000\,km of relay. In the 5G CATS cell, an uplink slot comes every 1.25\,ms, equivalent to 125\,km. A 20-round control message trace of the 5G CATS cell attributes 2.27\,ms of its 3.65\,ms median to the scheduling-request and grant handshake, which configured grants would remove.

\begin{observationbox}
\textbf{O1.} 5G CATS narrows the honest round trip more than tenfold. A simple use of default 5G fails because its uplink schedule, not the distance to the tracee, sets the round trip.
\end{observationbox}

\begin{table}[t]
\centering
\caption{Round-Trip Comparison of PowerCATS, Default 5G and 5G CATS}
\label{tab:urllc-embb}
\begin{tabular*}{\columnwidth}{@{\extracolsep{\fill}}lrrr@{}}
\toprule
\textbf{Metric} & \textbf{PowerCATS} & \textbf{Default 5G} & \textbf{5G CATS} \\
\midrule
Median round trip [ms] & -- & 40.25 & 3.92 \\
Minimum ($T_{\mathrm{min}}$) [ms] & -- & 17.11 & 3.11 \\
Threshold $T$ (p99.9) [ms] & 10.6--14.2 & 56.40 & 5.72 \\
Spread ($T - T_{\mathrm{min}}$) [ms] & -- & 39.28 & 2.60 \\
Estimated $d_\text{hide}$ [km] & -- & 3,928 & 260 \\
All relayed rounds rejected [km] & -- & 2,000 & 300 \\
Safe distance [km] & 291--619 & Never & 1,410 \\
\bottomrule
\multicolumn{4}{@{}p{\columnwidth}@{}}{\footnotesize *PowerCATS's threshold is its published one-way delay from Fig.~6(a) at zero distance, which we double as our conversion to a round trip. Its model includes an error allowance and has no measured distribution, so it has no comparable spread. Safe distance is the largest tracer-tracee distance at which a 3,000\,km attacker is still detected (Section~\ref{sec:eval-distance}).}
\end{tabular*}
\end{table}

\begin{figure}[t]
  \includegraphics[width=\columnwidth]{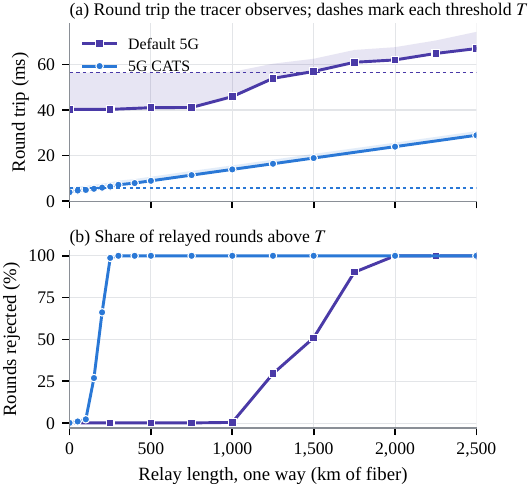}
  \caption{A fiber relay between the chip and its serving cell; 20 cell realizations $\times$ 1,000 recertifications per point. (a) Round trip the tracer observes, median and 99.9th-percentile band, with each threshold $T$ dashed. (b) Share of relayed rounds above $T$.}
  \label{fig:relay}
\end{figure}

\subsection{Q2: Detecting a Fiber Relay}
An attacker who moves a tracee into a restricted region must relay its traffic over fiber to an approved gNB, so the tracer receives valid replies that arrive late. Q2 asks whether the tracer can detect that delay. We sweep the relay from 0 to 2,500\,km, with 20 cell realizations and 1,000 recertifications per point. Figure~\ref{fig:relay} reports the median and 99.9th percentile of the round trip the tracer observes and the share of rounds above each system's threshold, and Table~\ref{tab:fiber-relay} lists selected points.

Under 5G CATS, the median round trip tracks the fiber-propagation line $2d/c_f$. It grows from 3.9\,ms at $d = 0$ to 28.9\,ms at $d = 2{,}500$\,km, a slope of 10.0\,\textmu s per kilometer of relay, and sits within 0.25\,ms of the fiber line at every point in the sweep. The tracer rejects 1\% of rounds at 50\,km, 27\% at 150\,km, 99\% at 250\,km, and all of them from 300\,km.

This is not the case under default 5G. The relay adds less than 1\,ms to the median through 750\,km and 5.6\,ms at 1,000\,km, where the fiber alone would add 10\,ms. The relay delay falls before the reply's wait for an uplink slot, so a later reply often still catches the same slot and the wait absorbs the delay. The tracer rejects 0.4\% of rounds at 1,000\,km, 51\% at 1,500\,km, and all of them only from 2,000\,km.

The 5G CATS median stays 32 to 38\,ms below default 5G's at every $d$ in Table~\ref{tab:fiber-relay}. From the median round trip, a 5G CATS tracer can tell a 50\,km change from a 100\,km change. A default 5G tracer cannot, because the frame absorbs delays of this size. PowerCATS publishes no relay sweep to compare against.

\begin{observationbox}
\textbf{O2.} A 5G CATS tracer rejects every round of a relay from 300\,km. Under default 5G, a relay of up to 1,000\,km is almost never rejected, and every round is rejected only from 2,000\,km.
\end{observationbox}

\subsection{Q3: Distance from Tracer to Tracee}
\label{sec:eval-distance}
A tracer may sit far from the tracee it checks. The farther it sits, the longer an honest reply takes, until an attacker in a restricted region can answer as fast as an honest tracee and the tracer can no longer tell them apart. Q3 measures how far a tracer can sit from a tracee and still detect that attacker, on the axes of PowerCATS~\cite{powercats} Fig.~6 against the same attackers.

\begin{figure*}[t]
  \includegraphics[width=\textwidth]{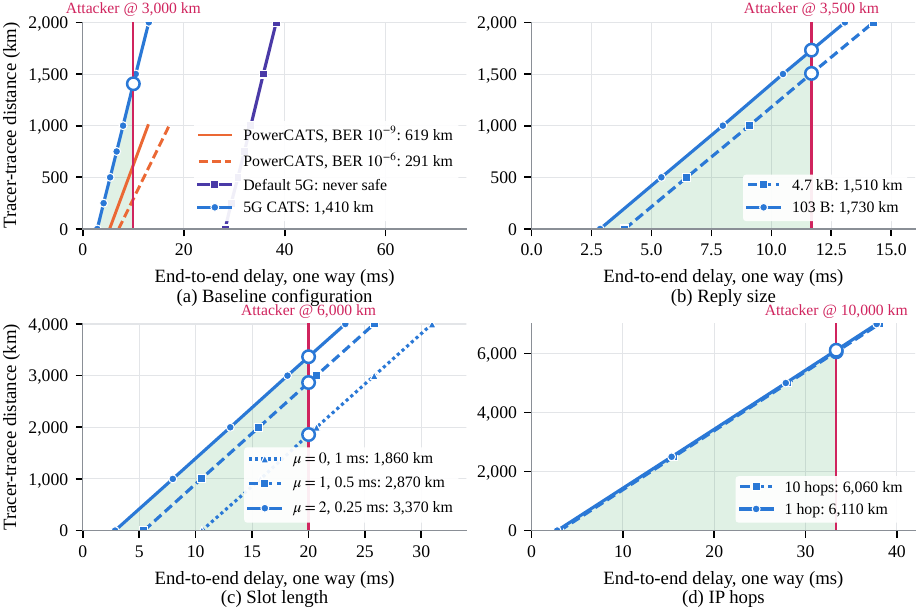}
  \caption{Tracer-tracee distance against end-to-end delay, the 5G counterpart of PowerCATS~\cite{powercats} Fig.~6. Lines show the 99.9th-percentile round trip, halved, pooled over 20 cell realizations. Rings mark where a line crosses the attacker, which is the largest safe distance, interpolated between measured distances. PowerCATS's published curves end at 1,000\,km. (a) Baseline, with PowerCATS's published curves. (b) Reply size. (c) Slot length. (d) IP hops.}
  \label{fig:distance}
\end{figure*}

Figure~\ref{fig:distance} plots the 99.9th-percentile round trip, halved to PowerCATS's one-way axis, against the fiber distance between the tracer and the 5G core. The vertical line is an ideal attacker whose reply travels at the speed of light with no routers, errors or queues. A tracee is safe wherever its line lies to the left of the attacker. Table~\ref{tab:safe-distance} gives the largest safe distance in each case. Against the 3,000\,km attacker, 5G CATS stays safe to 1,410\,km and PowerCATS's published curves to 291-619\,km. Default 5G is never safe since its 56.40\,ms threshold already exceeds the attacker's 20.0\,ms round trip at zero distance.

In the baseline panel, 5G CATS gains for two reasons. Its line starts lower, at 2.9\,ms one way against 5.3 to 7.1\,ms, and it rises more slowly, at 5.1\,\textmu s/km of fiber against 7.6 to 10.0\,\textmu s/km on the power line. PowerCATS's faster 400\,kHz band starts lower, at 1.6\,ms, but its published curve still reaches only 690\,km against the 6,000\,km attacker, where 5G CATS reaches 3,370\,km. Reply size, slot length and hops mainly shift the starting point: growing the reply from 103\,B to 4.7\,kB costs 220\,km, moving from $\mu = 2$ to $\mu = 0$ costs 1,510\,km, and moving from one to ten hops costs 105\,km. PowerCATS's curves are closed-form results with no queuing, while ours are simulated with scheduling, so the two are not measured the same way.

\begin{table}[t]
\centering
\caption{Largest Safe Tracer-Tracee Distance [km]}
\label{tab:safe-distance}
\begin{tabular*}{\columnwidth}{@{\extracolsep{\fill}}lrrr@{}}
\toprule
\textbf{Factor (attacker)} & \textbf{PowerCATS} & \textbf{Default 5G} & \textbf{5G CATS} \\
\midrule
Baseline (3,000\,km) & 291--619 & Never & 1,410 \\
Packet size (3,500\,km) & 195--910 & Never & 1,510--1,730 \\
Link speed (6,000\,km) & 390--690 & Never & 1,860--3,370 \\
Routing devices (10,000\,km) & 1,110--3,880 & 1,020 & 6,060--6,110 \\
\bottomrule
\multicolumn{4}{@{}p{\columnwidth}@{}}{\footnotesize *PowerCATS values are read from its published Fig.~6, and its ranges and those of 5G CATS span the settings in each panel. Default 5G uses its baseline configuration in every row.}
\end{tabular*}
\end{table}

\begin{observationbox}
    \textbf{O3.} Against a 3,000\,km attacker, a tracer can sit 1,410\,km from a 5G CATS tracee, more than twice PowerCATS's best published distance of 619\,km. Default 5G fails against this attacker at any distance.
\end{observationbox}

\subsection{Q4: Deployment Parameters}
\begin{figure}[t]
  \includegraphics[width=\columnwidth]{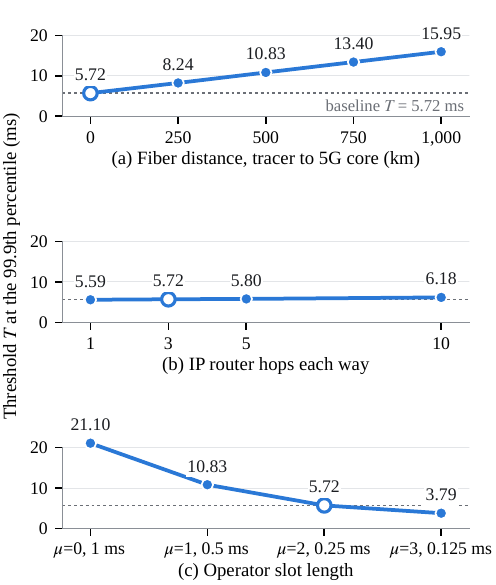}
  \caption{5G CATS threshold $T$ (99.9th percentile) as one deployment parameter changes. (a) Tracer-to-core fiber distance. (b) IP router hops on the same path. (c) Operator slot configuration; $\mu = 3$ is measured at 100 m instead of 300 m because FR2 mmWave range is shorter. All points pool 20 cell realizations, and the open circle marks the baseline.}
  \label{fig:sensitivity}
\end{figure}

A real tracer will not always sit in the position that gives the 5.72\,ms baseline, beside the 5G core on a clean cell with three IP router hops and $\mu = 2$. Q4 asks how much an attacker gains as the deployment moves away from that baseline. We sweep three parameters one at a time, and Figure~\ref{fig:sensitivity} shows their effect on the 5G CATS threshold $T$.

The backhaul and hops raise $T$ predictably. 1,000\,km of fiber between the tracer and the 5G core raises $T$ to 15.95\,ms (an added 10.25\,\textmu s per kilometer, mostly fiber plus the extra router hop every 300\,km). Ten IP hops instead of three raise $T$ to 6.18\,ms (65\,\textmu s per hop). Both move the whole distribution, so the spread stays at 2.62\,ms and 2.67\,ms against the baseline 2.60\,ms, and the estimated hidden relay barely changes. The tracer only has to add the known delay to $T$.

The operator's spectrum choice is the sharpest edge. The 5G CATS result depends on the slot configuration the operator provisions. At $\mu = 0$ (15\,kHz subcarriers, 1\,ms slot) $T$ sits at 21.10\,ms; at $\mu = 1$ (0.5\,ms slot) it is 10.83\,ms; at the $\mu = 2$ baseline (0.25\,ms slot) it drops to 5.72\,ms; and at $\mu = 3$ it reaches 3.79\,ms, although the shorter mmWave range places the chip 100\,m from the gNB. Unlike distance, slot length changes the spread, which gives estimated hidden relays of 973, 499, 260 and 165\,km. The chip and the tracer cannot recover a tight $T$ on top of a $\mu = 0$ carrier.

\begin{observationbox}
    \textbf{O4.} A tracer can budget for distance and router hops in advance, but the carrier's slot length sets how much relay an attacker can hide.
\end{observationbox}

\begin{table}[t]
\centering
\caption{Median Round Trip and Rejected Rounds Under a Passive Fiber Relay}
\label{tab:fiber-relay}
\begin{tabular*}{\columnwidth}{@{\extracolsep{\fill}}lrrrrr@{}}
\toprule
& \multicolumn{5}{c}{\textbf{One-Way Relay Distance $d$ [km]}} \\
\cmidrule{2-6}
\textbf{Profile / Metric} & \textbf{0} & \textbf{250} & \textbf{500} & \textbf{1,000} & \textbf{2,000} \\
\midrule
$2d/c_f$ (fiber delay) [ms] & 0.0 & 2.5 & 5.0 & 10.0 & 20.0 \\
Default 5G median [ms] & 40.3 & 40.3 & 41.0 & 45.9 & 62.0 \\
5G CATS median [ms] & 3.9 & 6.4 & 8.9 & 13.9 & 23.9 \\
Default 5G rejected [\%] & 0.1 & 0.1 & 0.1 & 0.4 & 100 \\
5G CATS rejected [\%] & 0.1 & 98.8 & 100 & 100 & 100 \\
\bottomrule
\multicolumn{6}{@{}l}{\footnotesize *$2d/c_f$ is the pure round-trip fiber propagation delay ($c_f \approx \tfrac{2}{3}c$).}
\end{tabular*}
\end{table}

\subsection{Q5: False Rejections of Honest Tracees}
\label{sec:eval-lease}
A tracer that sets its threshold at a percentile will reject a small share of honest rounds. If one rejected round erases the key, an honest tracee can be disabled by chance. If a rejection has no lasting effect, an attacker can keep computing between failures. Q5 asks whether 5G CATS avoids both outcomes.

We replay the round-by-round traces of the relay sweep under each rule, over 20 cell realizations of 1,000 rounds. This is an offline replay of measured round trips, not a simulated implementation of the lease, and the thresholds are fitted to the same rounds. A permanent lock fires on 9 of 20 honest chips under default 5G and on 11 of 20 under the 5G CATS radio profile. With execution leases, no honest chip is halted for more than one round, and the sequential test flags none of the 20. Against a relay on the 5G CATS profile, the test flags 14 of 20 chips at 50\,km, 20 of 20 at 100\,km with a median of 374 rounds, and 20 of 20 from 300\,km within 9 rounds.

The interlock fires this early because a 99.9th-percentile threshold is exceeded about once per thousand rounds, and a single excess is enough to end the chip. The honest misses show no clustering, with no chip above 4 misses, but twenty runs cannot validate an operational false-alarm rate. The test is weakest at 50\,km because the miss rate there, 0.95\%, sits just above the 0.9\% density the test flags, so runs longer than our 1,000 rounds would flag more chips. 

\begin{observationbox}
    \textbf{O5.} A permanent lock disables honest chips within 1,000 rounds, so the simple approach fails even when the radio is fast. In replay, execution leases keep every honest chip running, and the sequential test still flags every relayed chip from 100\,km.
\end{observationbox}

\section{Limitations}
Our results are conditional on a simulated cell. URLLC is not a default in the simulator, so we approximate it. In practice, it is a service category an operator provisions on top of a dedicated slice with mini-slot scheduling, configured-grant uplink, QoS 5QI 82, capability-2 UEs, conservative modulation with low-latency pre-emption, and typically a short TDD pattern like DDDSU. Public cellular networks almost always run default 5G and URLLC deployments at scale are limited to private industrial networks and a small set of operator pilots. This matches our target: approved sites are data centers and enterprise sites that can host private URLLC cell, and 5G CATS is not meant for consumer GPUs in homes.

Our ns-3 / 5G-LENA stack approximates such a slice with the parameters the simulator exposes: 0.25 ms slot at $\mu = 2$, an uplink opportunity every 1.25 ms, a $10^{-5}$ BLER target, UE processing capability 2, at most one HARQ retransmission, and RLC unacknowledged mode. The two most latency-sensitive URLLC primitives are not in 5G-LENA: configured-grant uplink and sub-slot mini-slots. In both profiles the chip's reply therefore still waits through a scheduling-request / grant handshake and lands on a full-slot boundary. These omissions should lengthen our round trip, while ideal beamforming, ideal RRC signaling and a single line-of-sight device on an unloaded cell should shorten them. We therefore do not claim that our numbers bound a real URLLC cell in either direction, and loaded and non-line-of-sight cells remain to be evaluated. 

Our default 5G profile is one eMBB configuration; a 30\,kHz cell would have a shorter TDD period and a shorter round trip. Our relay adds fiber delay above the radio link, so the cell's timing advance stays normal, and a radio-layer relay would change it. We assume that the tracer knows $T$ for a cell, and how to calibrate it without interference from the chip's owner is open.

We also simulate one device per cell. Our design assumes that the tracer staggers challenges across chips and that the cell reserves configured-grant resources for the CATS flow, so that recertification rarely preempts other traffic. How many chips one cell can recertify at a given round rate, and how often preemption would still be needed at hyperscale sites with thousands of chips, remain to be evaluated. The carrier binding is also unsimulated: the NEF subscription keeps the cell lookup outside the measured round trip, but a delayed or missed notification would postpone the detection of a cell change.

\section{Discussion and Related Work}
Ping-based location verification is the state of the art. The U.S. Chip Security Act (H.R. 3447 \cite{hr3447_chip_security_act}) would require location verification using feasible and appropriate techniques. The CSET assessment places such schemes alongside physical inspection as the most feasible suggested combination~\cite{cset_tracking_chips}. Internet landmark schemes and NVIDIA's opt-in fleet telemetry produce just a report of whether the chip is inside or outside the authorized region and hand the enforcement step off to the exporter, regulator or physical inspection. The 5G CATS specialty is that the enforcement step is built into the core of the system with the hardware CATS chip. PowerCATS~\cite{powercats} is the closest prior work, and its use of the power line exposes its timing to the channel noise and routing conditions of the grid.

A parallel track attests who a chip is rather than where. Tee and Happel~\cite{tee2026hardware} replace the on-chip identity with a hardware fingerprint derived from measurable chip variation. NVIDIA Confidential Computing on Hopper and Blackwell attests firmware state and CC-enabled mode through the NRAS (NVIDIA Remote Attestation Service) verifier but says nothing about physical location. Classical PUFs and supply-chain tracking behave similarly. All are complementary to 5G CATS: they attest, while 5G CATS supplies the location bound once identity is established and provides primitives to take action on non-compliant devices.

\section{Conclusion}
We designed 5G CATS, a chip-location certification scheme over a 5G URLLC slice with the radio co-packaged on the GPU, and evaluated its timing in simulation on an unloaded cell. A simple port to default 5G fails: its uplink schedule leaves a 39.28\,ms spread and absorbs relays of up to 1,000\,km, and in replay a permanent key erasure disables honest chips. At the baseline deployment the 5G CATS round trip is 5.72\,ms at the 99.9th percentile, and default 5G gives 56.40\,ms, almost ten times higher. A fiber-relay sweep reproduces the 10\,\textmu s/km propagation slope under 5G CATS, and the tracer rejects every round from 300\,km, against 2,000\,km under default 5G. Against a 3,000\,km attacker, a tracer stays safe to 1,410\,km, against at most 619\,km for PowerCATS's published model. Fiber backhaul and IP hops raise $T$ predictably and can be budgeted, but the operator's slot length sets the hideable distance, so results depend on the carrier provisioning short slots.

\section{AI Use}
We would like to acknowledge Anthropic's Claude, Google's Gemini and Perplexity AI's Perplexity for editing help, assistance in building the simulation and literature search. We certify that the content of this paper is accurate to the best of our knowledge and that all text was authored by the listed authors.

\bibliographystyle{IEEEtran}
\newpage
\bibliography{refs}
\end{document}